\documentclass[prl,twocolumn]{revtex4-1}

\usepackage{amsmath}
\usepackage{amssymb}
\usepackage{amsthm}

\usepackage{graphicx}
\usepackage{graphics}
\usepackage{bm}
\usepackage{hyperref}
\usepackage{epsfig}
\usepackage{dcolumn}

\newcommand\vex[1]{\mathbf{#1}}

\def\re{\mathrm{Re}}

\begin{document}


\title{Proposed Aharonov-Casher interference measurement of non-Abelian vortices in chiral p-wave superconductors}

\author{Eytan Grosfeld}
\affiliation{Department of Physics, University of Illinois,
1110 W. Green St., Urbana IL 61801-3080, U.S.A.}

\author{Babak Seradjeh}
\affiliation{Department of Physics, University of Illinois,
1110 W. Green St., Urbana IL 61801-3080, U.S.A.}

\author{Smitha Vishveshwara}
\affiliation{Department of Physics, University of Illinois,
1110 W. Green St., Urbana IL 61801-3080, U.S.A.}

\begin{abstract}

We propose a two-path vortex interferometry experiment based on the Aharonov-Casher effect for detecting the non-Abelian nature of vortices in a chiral p-wave superconductor. The effect is based on observing vortex interference patterns upon enclosing a finite charge of externally controllable magnitude within the interference path. We predict that when the interfering vortices enclose an odd number of identical vortices in their path, the interference pattern disappears only for non-Abelian vortices. When pairing involves two distinct spin species, we derive the mutual statistics between half quantum and full quantum vortices and show that, remarkably,  our predictions still hold for the situation of a full quantum vortex enclosing a half quantum vortex in its path.  We discuss the experimentally relevant conditions under which these effects can be observed.
\end{abstract}

\maketitle


\emph{Introduction.}---It has been long understood theoretically that topology in two spatial dimensions (2d) permits anyonic exchange statistics of identical particles in addition to the bosonic and fermionic possibilities~\cite{LeiMyr77a}. Non-Abelian anyons, for which different exchange operations may not commute, have attracted great attention, particularly as building blocks of a topologically protected quantum computer~\cite{Kit03a}. Over the past several years, the quest for realizing non-Abelian anyons has been focused on the fractional quantum Hall (FQH) system (e.g. at filling factor $5/2$) and the two-dimensional chiral p-wave superconductor (CpSC)~\cite{ReaGre00a}.
Theoretical proposals for detecting non-Abelian anyons in FQH systems have exploited quantum interference~\cite{SteHal06a}, leading to active experimental study.  In contrast, analogous explorations in the latter system have been scarce~\cite{DasNayTew06a}. In this Letter we propose an interferometry experiment for directly detecting non-Abelian anyons in a CpSC.

In 5/2 FQH proposals, an Aharonov-Bohm interference pattern is expected when $e/4$ quasiparticles take two paths that enclose a region of flux and an even number of $e/4$ quasiparticles. Their non-Abelian statistics manifests itself as the effacing of this pattern when another $e/4$ quasiparticle is introduced in this region. Extending this idea to the setting of a CpSC poses a conceptual challenge. Non-Abelian anyons in a CpSC are neutral Majorana modes bound to vortex cores~\cite{KopSal91a}, requiring an alternative means of interference.  We show that the key to resolving this issue lies in the lesser known Aharonov-Casher (AC) effect~\cite{AhaCas84a}, which is an elegant `dual' to the Aharonov-Bohm effect in that the roles of charge and flux are interchanged~\footnote{We thank Dale Van Harlingen for bringing the AC effect in a superconductor setting to our attention.}. This is our first central result.

\begin{figure}[t]
\begin{centering}
\includegraphics[width=2.6in]{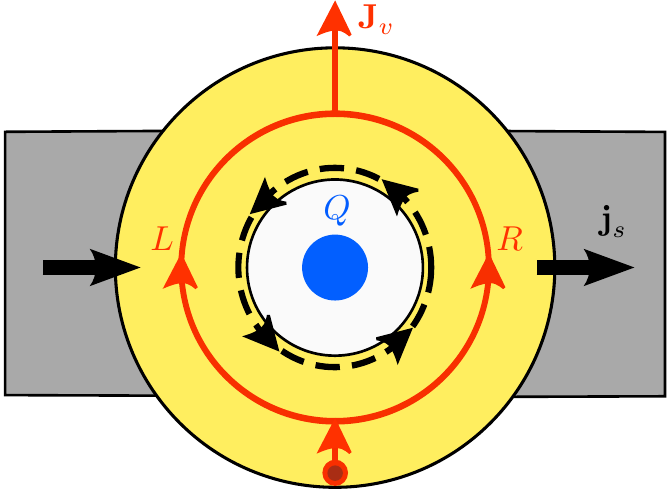}
\end{centering}
\caption{(color online) Geometry of the proposed setup. A mesoscopic CpSC ring (yellow) is connected to leads (grey). A super-current density $\vex j_s$ is maintained in the sample. Due to the Magnus force, vortices (red) flow from bottom to top and choose one of two paths ($L$ and $R$) to circumvent the central hole. A measurable voltage is induced across the leads by the vortex current $\vex J_v$ via the Josephson relation. Charge $Q$ on the central island (blue) is controlled by gates (not shown). A vortex may be stabilized at the hole, inducing currents around the hole edges shown by dashed arrows.}
\label{fig:setup}
\vspace{-0.5cm}
\end{figure}

The simplest incarnation of a CpSC,  the `one-component CpSC,' has been recently proposed in the superconducting state on the surface of a strong topological insulator induced by proximity to a conventional superconductor~\cite{FuKane08}, where locking of spin and orbital degrees of freedom effectively ensures that only a single component participates in the condensate. The situation is more complex in a `two-component CpSC' with an unconventional pairing similar to that of superfluid $^3$He-A~\cite{Leg75a} due to the existence of two distinct spin components. A prominent solid-state candidate in this class is the perovskite structured material Sr$_2$RuO$_4$ (SRO)~\cite{MacMae03a}. The presence of both components doubles the number of Majorana states in the core of a full quantum vortex (FQV) and renders their exchange statistics trivial. Fortunately, the order parameter allows for a topologically stable half quantum vortex (HQV) with a single Majorana bound state. A HQV carries a quantized flux of $h/4e$ and acts as a vortex in just one spin component. It has been argued that certain bounded geometries may favor HQVs energetically~\cite{ChuBluKim07a}. Recent tour de force experiments in Budakian's lab at UIUC have found strong evidence for the presence of $h/4e$ vortices bound to the inner radii of mesoscopic  SRO rings~\cite{Raffi10}. Inspired by these experiments, we tailor our proposal to ring geometries and consider the effects of two spin components. It is likely that, due to stability conditions, itinerant vortices are FQVs. We show that, quite remarkably, under some reasonable conditions much of our proposed interference physics is preserved even when FQVs go around a HQV stabilized around a mesoscopic hole. This is our second central result.

\emph{Proposed setup and mechanism.}---The geometry that we propose is shown in  Fig. \ref{fig:setup}. An annulus formed by a mesoscopic CpSC sample is connected to leads. A supercurrent density, $\vex j_s$, is maintained in the sample between the leads. Within the hole, an island supports variable charge, $Q$, whose magnitude is controlled by gating. An applied magnetic field can induce quantized flux through the hole. The interferometry is brought about by a conspiracy of several factors as described below.

The supercurrent induces a transverse force on vortices, which in its simplest Magnus form  is proportional to  $\vex j_s\times\Phi_V \hat{\vex z}$~\cite{Tho99},
 where $\Phi_V \equiv h/e^*$ is the flux
quantum carried by the vortex and $\hat{\vex z}$ is the unit vector normal to the superconducting plane. In a one-component CpSC, $e^*=2e$. In a two-component CpSC, $e^*=2e$ ($4e$) for a FQV (HQV). This force causes vortices that enter from the bottom of  the sample to move toward the top edge. These vortices can circumvent the central hole of the annulus by one of two paths, 
with amplitudes $t_L$ and $t_R$ and associated partial waves $|\psi_L\rangle$ and $|\psi_R\rangle$ denoting the internal states of the vortices in the system. In principle, vortices
can be guided around the annulus by  etching paths in the superconductor that would pin vortex trajectories.
The associated vortex current is
\begin{eqnarray}\label{eq:Jv}
	J_v\propto |t_L|^2+|t_R|^2+2 |t_L| |t_R| \re\left[e^{i\varphi}\langle \psi_L|\psi_R\rangle\right],
\end{eqnarray}
with $\varphi=\arg(t_L^*t_R)$. The last term represents the interference between the two paths. As vortices move across the sample, they generate a superconducting
phase difference normal to their motion. By the Josephson relation, a voltage drop $V_s\propto J_v$ develops along $\vex j_s$. A measurement of $V_s$, or equivalently, of differential resistance for fixed current, therefore constitutes an interferometry of the two vortex paths.

As a tuning parameter for the interference pattern, we propose the contribution to the phase $\varphi$ stemming from the Aharonov-Casher effect, namely,
the geometric phase picked up by the itinerant vortices upon encircling the charge $Q$ on the central island. This non-local effect entails a magnetic 
moment $\boldsymbol{\mu}$ moving in a static external electric field $\vex{E}$. The adiabatic transport of such a moment around a closed loop causes it 
to accumulate a quantum phase~\cite{AhaCas84a,AhaRez00} 
$
\phi_{AC}=\oint\boldsymbol{\mu}\times\vex{E}\cdot d\vex{l}/({\hbar c^2}).
$
A fluxline carries a moment density $\epsilon_0 c^2\Phi_V$ along its length. 
When it encircles a total charge $Q$ (of any distribution), it acquires a geometric phase $\phi_{AC}=\Phi_VQ/\hbar=2\pi Q/e^*$~\cite{suppl}. It can be shown that the phase $\phi_{AC}$ is truly non-local in that this form remains unaltered even in the case of an external charge placed at the center of a superconducting ring inspite
of screening effects~\cite{AhaCas84a,AoZhu95a}.  
Consequently,
as a function of charge $Q$, $J_v$ in Eq.~(\ref{eq:Jv}) and so the voltage $V_s$ would exhibit oscillations with period $e^*$.

This picture changes drastically when a flux is applied through the superconductor so as to nucleate a non-Abelian vortex on the hole in the center of the sample. 
Now the interference term also contains a braiding operation of the itinerant vortices around this vortex. As discussed in previous work~\cite{Iva01a,DasNayTew06a}, when the itinerant vortices 
are themselves non-Abelian, this braiding yields $\langle \psi_L|\psi_R\rangle=0$; the interference term in Eq.~(\ref{eq:Jv}) vanishes and $V_s$  becomes independent of $Q$. Analogous to the
FQH case, we propose the obliteration of the Aharonov-Casher oscillations as the clear signature of non-Abelian statistics of vortices.

For a one-component CpSC, a vortex carries the smallest flux quantum and these considerations suffice. However, a FQV in a two-component CpSC contains two Majorana modes. 
One might therefore question whether this interferometry would work if the itinerant vortices are FQVs. Furthermore, the coupling between these two modes would split their energies and combine them into a regular fermion. We now consider this situation in more detail and show that there is an experimentally relevant window of parameters for which our proposed interferometry does still work when FQVs go around a HQV.

\emph{Majorana modes in a two-component CpSC.}---The Hamiltonian for the two-component 2d CpSC is
$
H=\left(\begin{array}{cc}
h & \Delta\\
\Delta^\dag & -h^T
\end{array}\right),
$
where $h=(\frac{p^2}{2m}-\mu)I$ is the single particle hamiltonian ($\mu$ is the chemical potential and $I$ is the identity matrix), and
\begin{eqnarray}
	\label{eq:pairing-form} \Delta=\frac{i}{2}v_\Delta e^{i\Phi/2}\{\partial_x-i\partial_y,({\boldsymbol \sigma}\cdot {\mathbf d})\sigma_2\}e^{i\Phi/2},
\end{eqnarray}
is the pairing term. Here ${\boldsymbol\sigma}=(\sigma_{1},\sigma_{2},\sigma_{3})$
are the Pauli matrices, $\vex d$ defines the
direction of pairing in the space spanned by the spin-triplet states,
$\Phi$ is the phase of the order parameter, and $v_\Delta$ is a constant gap velocity. 
We represent the electronic operators as a Nambu spinor, $\Psi^\dag=(\psi_\uparrow^\dag, \psi_\downarrow^\dag,\psi_\uparrow, \psi_\downarrow)$,
where $\psi_\sigma^\dag$ creates an electron with spin $\sigma$. Expanded in quasi-particle modes of energy $E$ we have $\Psi = \sum_{E}\chi_{E}\Gamma_{E}$, where $\chi_E^T=(u_{E\uparrow}, u_{E\downarrow}, v_{E\uparrow}, v_{E\downarrow})$ is the wavefunction of the Bogoliubov quasiparticle destroyed by $\Gamma_E$.
Particle-hole symmetry dictates that $\Gamma_{E}=\Gamma_{-E}^{\dag}$ and thus $u_{E\sigma}=v_{-E\sigma}^{*}$.
At zero energy, $\Gamma_{0}=\Gamma_{0}^{\dag}\equiv\gamma$ would be a Majorana fermion. 

The Hamiltonian respects several more symmetries in the absence of external fields. It is invariant under a global SU$(2)$ transformation
$
\Psi = \left(\begin{array}{cc}
S & 0\\
0 & S^{*}\end{array}\right)\Psi'
$,
$\vex d = R^{-1}_S \vex d'$, where $S\in$~SU(2) and $R_S$ is the corresponding rotation~$\in$ SO(3).
The Hamiltonian is also invariant under a rotation by angle $\phi$ around the z-axis, compensated by $\Phi\to\Phi+\phi$, and a 
suitable rotation of $\vex d$. Finally, there exists a $\mathbb{Z}_2$ symmetry under the mapping ${\mathbf d}\to -{\mathbf d}$ and $\Phi\to \Phi+\pi$. 

It is this last $\mathbb{Z}_2$ symmetry which allows the presence of half-quantum vortices. When encircling a HQV, the phase of the order parameter changes by $\pi$. The extra angular momentum needed to produce a single-valued Cooper pair wavefunction is provided by rotating $\vex d$ to $-\vex d$. We consider a configuration where $\vex d$ rotates in an ``easy'' plane normal to a unit vector $\vex{\hat n}$, being the maximal symmetry breaking that still allows a HQV. Applying a magnetic field will explicitly break the SU(2) symmetry. We focus on the SU(2) symmetric case so as to freely choose this axis without affecting the result.

As the vector field $\vex d$ in the presence of a FQV does not contain any winding, the two zero energy Majorana modes can be found by locally rotating $\vex d$ to the y-axis, for which the Hamiltonian decouples into spin components, see Eq. (\ref{eq:pairing-form}). The two Majorana modes are found to be 
\begin{widetext}
\begin{equation}
\left(\begin{array}{c}
\gamma_{1}\\
\gamma_{2}\end{array}\right)=\frac{1}{\sqrt{2}}\sum\left[e^{i\tilde\Phi/2}f S_{\vex d}
\left(\begin{array}{c}
\psi_{\uparrow}\\
\psi_{\downarrow}
\end{array}\right)
+e^{-i\tilde\Phi/2}f^{\ast}S_{\vex d}^{\ast}
\left(\begin{array}{c}
\psi_{\uparrow}^{\dagger}\\
\psi_{\downarrow}^{\dagger}\end{array}\right)
\right], \label{eq:majorana}\end{equation}
\end{widetext}
where $S_{\vex d}=e^{-i \delta \boldsymbol{\sigma}\cdot\vex{\hat n}/2}$ with $\hat{\vex n}$ the unit normal to the plane of $\vex d$ and $\hat{\vex y}$, $\cos\delta=\vex d\cdot\vex{\hat y}$, $\tilde\Phi$ is the local phase of the order parameter in the absence of the vortex, and $f(r)$ is an exponentially-decaying function of the distance from the vortex, determined by the vortex profile of $\Delta$~\cite{GurRad07}.

For a HQV, when $\vex d$ rotates normal to $\vex{\hat n}$ we may choose the y-axis at the intersection of the plane of rotation of $\vex d$ and xy-plane, so that $\vex{\hat n}=(\sin\beta,0,\cos\beta)$. Then the single Majorana mode of the HQV, $\gamma'$, is also given by Eq.~(\ref{eq:majorana}) in one of the spin components, with $S_{\vex  d}=e^{-i \beta \boldsymbol{\sigma}\cdot\vex{\hat y}/2}$, which takes $\vex{\hat n}\mapsto\vex{\hat z}$~\cite{suppl}.

Now consider a HQV on the hole in Fig.~\ref{fig:setup} and a FQV in the CpSC with a $\vex d$-vector determined by the local $\vex d$ prescribed by the HQV. The situation relevant to our proposal is of transporting the FQV around the HQV adiabatically. The Majorana mode at the HQV experiences a relative change of phase $\tilde\Phi$ by $2\pi$ as the FQV encircles it. Therefore, $\gamma'\to -\gamma'$. At the FQV, $\vex d$ rotates by angle $\pi$, i.e. $\delta\to\delta+\pi$. When $\beta=0$ ($\vex{\hat n} || \vex{\hat z}$), $S_{\vex  d}\to \exp\left(-i\frac{\pi}{2}\sigma_{3}\right)S_{\vex  d}=-i\sigma_{3}S_{\vex  d}$. Also, $\tilde\Phi\to\tilde\Phi+\pi$ so altogether the Majorana modes at the FQV transform as 
$
\gamma_1\to\gamma_1,\quad \gamma_2\to-\gamma_2.
$

In the general case $\beta\neq 0$, $S\to -i(\boldsymbol{\sigma}\cdot\vex{\hat n})S$. This is where our specific choice for $\vex{\hat n}$ pays off: we see that $\boldsymbol{\sigma}\cdot\vex{\hat n}= e^{-i\beta\sigma_2/2} \sigma_3 e^{i\beta\sigma_2/2}$, with $e^{i\beta\sigma_2/2}$ being a real-valued rotation, so that by rotating to a Majorana basis 
\begin{equation}
\left(\begin{array}{c}
\bar\gamma_{1}\\
\bar\gamma_{2}\end{array}\right)
=e^{i\frac{\beta}{2}\sigma_{2}}
\left(\begin{array}{c}
\gamma_{1}\\
\gamma_{2}
\end{array}\right), \label{eq:basis-change}
\end{equation} 
we find that altogether
\begin{equation}\label{eq:FQVexch}
\bar\gamma_1\to\bar\gamma_1,\quad \bar\gamma_2\to-\bar\gamma_2,\quad \gamma' \to -\gamma'.
\end{equation}

Thus, the statistics in the new Majorana basis is the same as when $\bar\gamma_2$ and $\gamma'$ are braided, while $\bar\gamma_1$ is unaffected.
These are the same operations as those for braiding two HQVs binding Majoranas $\gamma_2$ and $\gamma'$~\cite{Iva01a}. Similarly, therefore, we find $\langle \psi_L | \psi_R \rangle=0$.
 

While the above corresponds to the SU(2) invariant situation, symmetry-breaking terms such as a Zeeman term or spin-orbit interactions would split the degeneracy between 
the two Majorana modes on the FQV. For two such modes $\gamma_1$ and $\gamma_2$, having  wave-functions $\chi_1$ and $\chi_2$ respectively, the 
associated coupling term assumes the form
$
H_{1}=i\lambda\gamma_{1}\gamma_{2}.
$
It should be noted that this term is invariant under the transformation introduced in Eq.~(\ref{eq:basis-change}), suggesting a larger symmetry in the low energy sector. Among the terms of the single particle Hamiltonian that contribute to $\lambda$, the most relevant is the Zeeman coupling~\cite{suppl}, 
$
h_{Z}=-g\mu_B {\mathbf B}\cdot{\boldsymbol \sigma}/2,
$
which explicitly breaks the SU$(2)$ symmetry of the underlying Hamiltonian.
In Nambu space, the coupling takes the form
$
H_{1}=\left(\begin{array}{cc}
h_{Z} & 0\\
0 & -h_{Z}^{T}
\end{array}\right)\
$.
 In the leading perturbation theory~\cite{suppl} we find $\langle\chi_{2}|H_{1}|\chi_{2}\rangle  =  \langle\chi_{1}|H_{1}|\chi_{1}\rangle=0$, and
\begin{equation}
\lambda=-i\langle \chi_2|H_1|\chi_1\rangle=-g \mu_B {\mathbf B}\cdot {\mathbf d}/2.
\end{equation}
We see that when ${\mathbf B}\perp {\mathbf d}$ no coupling is generated. This is in fact an exact result: When $\vex d$ lies in the xy-plane, the effect of the Zeeman term is a relative shift of the chemical potential of the two Majorana modes. We therefore get different functions $f_{1,2}(r)$ associated with the two spinors $\chi_{1,2}$, but no shift in energy. When the Zeeman term dominates the energetics, it will render $d$ perpendicular to $B$, so that the two Majorana modes decouple. However, the combination of spin-orbit interaction and Zeeman term conspire to split the two Majorana modes.

\emph{Experiment.}---Piecing together the various components, we arrive at the following steps. The parentheses refer to a two-component CpSC. i) In the absence of an applied field through the hole in Fig. \ref{fig:setup}, drive a supercurrent $\vex j_s$; ii) Vary the charge $Q$ on the island enclosed by the hole. The differential resistance associated with the current ought to
show AC oscillations as a function of $Q$ due to vortex interference.  The period of oscillation would be $2e$ ($4e$ for itinerant HQVs); iii) Apply a field through the hole to nucleate a vortex (HQV). The interference pattern should disappear; iv) Increase the field to nucleate a second vortex (FQV). The interference pattern (for itinerant FQVs) should reappear. If the interference pattern is detected in ii) and does not disappear in step iii), the vortices either do not possess or cannot maintain their non-Abelian character over the time of the experiment.

\emph{Discussion.}---
Our proposal requires several conditions to be met.
For the AC effect, the vortices must be well separated from the charge on the island in Fig.~\ref{fig:setup}. So, the dimension $L$ of the hole must be greater than  the magnetic penetration depth. Our proposal requires the coherent quantum motion of vortices.
There is evidence that vortices can act as quantum objects~\cite{EliWacSoh93a,Hoo00a}. The biggest obstacle for such motion is dissipation in the vortex core. However, in the range of energies that we consider ($<$ minigap) the core dynamics is effectively frozen. Another impediment lies in overcoming vortex pinning by impurities. Fortunately, superconductivity in SRO requires some of the cleanest crystals, evidenced by the observation of a fairly regular vortex lattice~\cite{Rid98a}. In addition, methods have been developed using the concept of a flux flow transistor to create an ``easy'' channel for the flow of vortices, thus considerably reducing the critical current for the vortex flow. An alternative for achieving quantum behavior would be to replace Abrikosov vortices with vortices trapped in Josephson junction arrays as in the original experiment~\cite{EliWacSoh93a}.

Some conditions must be met for any statistical interaction to be observed. First, the system must maintain adiabaticity. Therefore, the rate for the motion of the vortices must be smaller than the smallest energy gap in the
problem. This is the minigap that separates the Majorana zero
mode in the vortex core and the next midgap state, $\hbar v_\Delta/L$. So, the velocity of itinerant vortices must satisfy $v\ll v_\Delta$. This is the most important condition for a one-component CpSC. 
Second, the Majorana modes must remain coherent during the braiding operations. In a two-component CpSC, the splitting $\lambda$ of the two Majorana modes on the FQV results in decoherence. So, in order to be sensitive to the braiding of Majorana modes in the presence of a HQV the braiding operation must run over a time shorter than $\pi\hbar/\lambda$, therefore $\lambda L/(\pi\hbar)<v\ll v_\Delta$.  

We now estimate the applicable range of parameters in SRO, where $m\sim10$ electron's mass,  $\hbar v_\Delta\sim10^{-3}$~eV\AA\ and $g\approx2.5$~\cite{MacMae03a}. The vortex velocity is related through the Josephson relation to the voltage drop $V_s$ across the superconductor: a
vortex crossing the sample changes the superconducting phase by $2\pi$. So, the voltage sensitivity $V_s$ when $N_{v}$ vortices move across the sample is given by $V_s/N_v\approx\pi\hbar v/(eL)\ll 10^{-7}$~volt for $L\approx 1$ micron.
The conditions for adiabaticity and coherence of FQV Majorana modes yield $\vex B\cdot \vex d\ll16$~G, which needs only to be satisfied locally over a coherence length at the FQV. 
In fact, the orientation of the $\vex d$-vector is still an unresolved issue; while a large enough Zeeman field along the z-axis ought to overcome spin-orbit interaction
to render $\vex d\perp\vex B$~\cite{MurIshKit04a,Raffi10}, recent studies show that this orthogonality might even be the natural configuration in SRO~\cite{YosMiy09a}. Additionally, the temperature must be smaller than the minigap not to excite other vortex core states which may lead to decoherence, i.e. $T < \hbar v_\Delta/L\sim 1$~mK. Other constraints on temperature might be imposed by the coherence condition discussed above. We believe that these conditions, while stringent, could still be realized in SRO.

Another issue in SRO is the existence of multiple layers. It is possible that this renders the exchange statistics of HQVs Abelian. (The non-Abelian nature of excitations in SRO has also recently been questioned in other ways~\cite{RagKapKiv10}.) Our proposal can be used to see if this is indeed the case. Multiple layers may also result in decoherence. The interlayer tunneling amplitude of Majorana fermions, $t$, depends on microscopic parameters (this is different from tunneling amplitude of Cooper pairs). Consequently, in the absence of experimental studies, it is not clear whether the limit of one independent Majorana mode per layer would hold or in fact the Majorana modes would completely hybridize to form a chiral low energy Majorana channel along the c-axis. If tunneling is present but small, the extra energy scale would introduce a decoherence time $\sim\hbar/t$ for our experiment~\cite{suppl}. 

\emph{Conclusion.}---The proposal advocated in this Letter presents a smoking-gun signature of non-Abelian statistics of vortices in a superconductor. Our proposal hinges on vortex interferometry based on the AC effect; a systematic experimental study of this effect in CpSC's would in itself be an important advancement. We have argued that under reasonable conditions in a two-componenent CpSC, not only HQVs but also FQVs encircling stationary HQVs would exhibit non-Abelian statistics due to the Majorana modes in the vortex core. Our predicted signature of such statistics in CpSC's is the clear suppression of the AC interference pattern in our proposed set of experiments. 



We extend our gratitude to R. Budakian and D. Van Harlingen for inspiring discussions and for sharing their unpublished results. This work has been supported by the ICMT at UIUC (E.G. and B.S.), the NSERC of Canada (B.S.) and the CAS fellowship at UIUC (S.V.) This material is based upon work supported by the U.S. Department of Energy, Division of Materials Sciences under Award No. DE-FG02-07ER46453, through the Frederick Seitz Materials Research Laboratory at the University of Illinois at Urbana-Champaign.


\section{Supplemental Material}
Here we discuss some details of the Aharonov-Casher effect, our derivation for the Majorana zero modes in half quantum and full quantum vortices in a two-component chiral p-wave superconductor, the splitting of Majorana modes in the full quantum vortex, and the effects of multiple layers in SRO.

\section{The Aharonov-Casher phase}

The Aharonov-Casher (AC) effect is closely related to the more well-known Aharonov-Bohm (AB)  effect. They are essentially the same effect, viewed from two relatively different reference frames and are both special cases of the more general Berry's phase. 

In the AB effect the quantum mechanical wave function of a charged particle acquires a geometric Berry's phase,
\begin{equation}
\phi_{AB} = \frac Q\hbar \oint \vex A\cdot d\vex l,
\end{equation}
where $\vex A$ is the vector potential and $d\vex l$ the element of the path of the particle. For a charge $Q$ circling a region of flux $\Phi = h / e^*$, we have $\phi_{AB}=2\pi Q/e^*$. This is nonzero even if the flux is completely confined to an area that has no overlap with the particle's trajectory.

In the AC effect, a moving fluxline acquires a Berry's phase as it circles a region of electric charge. This is the same AB phase as above but in the reference  frame where the charge is stationary. Using the Lorentz transformation of electromagnetism, the simplest form of the AC phase is~\cite{AhaCas84a}
\begin{equation}\label{eq:ACmuE}
\phi_{AC} = \frac 1{\hbar c^2} \oint \boldsymbol{\mu}\times\vex E\cdot d\vex l,
\end{equation}
when a particle carrying a magnetic moment $\boldsymbol\mu$ closes a loop in an electric field $\vex E$.

Let us explicitly show that this is the same as the AB phase for any shape of fluxline  and irrespective of the charge distribution, as long as the flux is confined. It was originally shown in Ref.~\onlinecite{AhaCas84a} that for a point charge $Q$ at $\vex r_0$ and a moment $\boldsymbol{\mu}$ at $\vex r$
\begin{equation}\label{eq:ACvp}
\vex E\times\boldsymbol{\mu} =  Q{c^2} \vex A_\mu,
\end{equation}
where $\vex A_\mu$ is the vector potential of the moment at $\vex r_0$, the position of the charge. A fluxline is a collection of infinitesimal magnetic moments $d\boldsymbol\mu$ along its length (with a linear density $d\boldsymbol{\mu} = \epsilon_0 c^2 \Phi d\vex s$ where $d\vex s$ is an element of length of the fluxline). Using Eq.~(\ref{eq:ACvp}) in Eq.~(\ref{eq:ACmuE}) we find quite generally
\begin{eqnarray}
\phi_{AC} 	&=& \frac Q{\hbar} \int ds \oint (d\vex A_\mu/ds)\cdot (-d\vex l) \\
		&=& \frac{Q}{\hbar} \int d\Phi \\
		&=& 2\pi Q/e^*.
\end{eqnarray}
We used the fact that $\oint d\vex A_\mu\cdot(-d\vex l)$ is nothing but the flux, $d\Phi$, due to the moment $d\boldsymbol\mu$ as seen by the charge if it were to go around in an inverted loop, but in the {\em same} direction as the fluxline. The integration over the fluxline then gives the total flux carried by the fluxline, since the flux is confined completely inside the loop. This result does not depend on the distance between the fluxline and point charge. So, it simply generalizes to any geometric charge distribution as long as flux is confined and has no overlap with the charge distribution.

\section{Majorana modes of HQV and FQV}

\subsection{Full quantum vortex}

Let us take the $\vex d$ vector to be constant, pointing at some arbitrary
direction. We perform a SU(2) transformation
\begin{equation}\label{eq:SU2}
\Psi=U\Psi',
\quad\quad
U=\left(\begin{array}{cc}
S & 0\\
0 & S^{*}\end{array}\right).
\end{equation}
This maps $\vex{d}\to \vex d'=R_S\vex d$, where $R_S$ is the SO(3) rotation corresponding to $S$, since
\begin{equation}
S^{\dagger}(\vex{\boldsymbol{\sigma}\cdot\vex{d}})\sigma_{2}S^{*}=S^{\dagger}(\boldsymbol{\sigma}\cdot\vex{d})S\sigma_{2} = \boldsymbol{\sigma}\cdot R_S\vex d\; \sigma_2.
\end{equation}
We choose $S$ to be a rotation that takes $\vex{d}$ to the y-axis,
\begin{equation}\label{eq:SS}
S=e^{i\frac{\delta}{2}(\boldsymbol{\sigma}\cdot\hat{\vex{n}})}
\end{equation}
where the unit vector $\hat{\vex n}$ is orthogonal to the plane of $\vex{d}$ and $\hat{\vex y}$ and $\delta=\cos^{-1}(\vex d\cdot\hat{\vex y})$. After this rotation, $\vex{d}\to\vex d'=(0,1,0)$.
The resulting Hamiltonian is now diagonal in spin indices,
\begin{widetext}
\begin{equation}
H'=\left(\begin{array}{cc}
\left(\frac{p^{2}}{2m}-\mu\right)I & iv_{\Delta}e^{i\theta/2}(\partial_{x}-i\partial_{y})e^{i\theta/2}I\\
iv_{\Delta}e^{-i\theta/2}(\partial_{x}+i\partial_{y})e^{-i\theta/2}I & -\left(\frac{p^{2}}{2m}-\mu\right)I\end{array}\right)
\end{equation}
\end{widetext}
so that it decouples into the two spin-components, and we get two
independent zero energy states (see next section for details). Here, for simplicity we also assumed a vanishing small vortex core so that $v_\Delta$ is constant everywhere but at the origin. Using the well known result for a HQV, we can write the two zero energy solutions for $H'$,
\begin{equation}
\chi'_{1}=\frac{1}{\sqrt{2}}\left(
\begin{array}{c}
f v_\uparrow\\
f^{*} v_\uparrow
\end{array}
\right),
\quad\quad
\chi'_{2}=\frac{1}{\sqrt{2}}\left(
\begin{array}{c}
f v_\downarrow \\
f^{*} v_\downarrow
\end{array}
\right)\label{eq:spinors}
\end{equation}
where $v_\uparrow=(1,0)^T$, $v_\downarrow=(0,1)^T$ and
\begin{equation}
f(r) = e^{-i\pi/4}e^{-mv_\Delta r}\begin{cases}
\begin{array}{c}
J_{0}(\kappa r)\\
I_{0}(\kappa r)\end{array} & \begin{array}{c}
\mu>\frac{1}{2}mv_{\Delta}^{2},\\
\mu<\frac{1}{2}mv_{\Delta}^{2},\end{array}\end{cases}
\end{equation}
and $\kappa\equiv \sqrt{2m|\mu-\frac12mv_\Delta^2|}$.
Rotating back to the original basis, the Majorana modes are given by Eq. (5) in the manuscript.

\subsection{Half quantum vortices}

\begin{figure}[b]
\begin{center}
\includegraphics[height=1.8in]{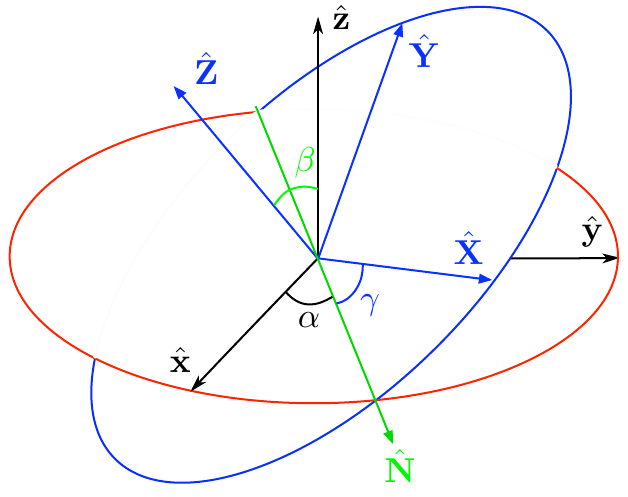}
\caption{The Euler angles and the plane of rotation of $\vex d$ for the HQV.}
\label{fig:euler}
\end{center}
\end{figure}

Let us now consider a $\vex d$ vector that rotates by $\pi$ in a tilted plane of rotation, as shown in Fig.~\ref{fig:euler}. We use the standard notation of the Euler angles, taking $\vex{d}$ to be along the $X$-axis, which sets it to be
\begin{widetext}
\begin{equation}
\vex{d}(\vex{r})=(\cos\alpha\cos\gamma-\cos\beta\sin\alpha\sin\gamma,\cos\gamma\sin\alpha+\cos\alpha\cos\beta\sin\gamma,\sin\beta\sin\gamma).
\end{equation}
Choosing $\gamma({\bf r})=\theta/2$, in the polar coordinates $\vex{r}=(r,\theta)$,
the $\vex{d}$ vector rotates by $\pi$ as we go around the origin,
\begin{equation}
\vex{d}(r,\theta)=\left(\cos\alpha\cos\frac{\theta}{2}-\cos\beta\sin\alpha\sin\frac{\theta}{2},\cos\frac{\theta}{2}\sin\alpha+\cos\alpha\cos\beta\sin\frac{\theta}{2},\sin\beta\sin\frac{\theta}{2}\right)
\end{equation}
\end{widetext}

The rotation
\begin{equation}
S=e^{i\frac{\beta}{2}\boldsymbol\sigma\cdot \hat{\vex N}}
\end{equation}
takes $\vex{d}\to\vex{d}'=R_S \vex d$ in the xy-plane. So,
\begin{equation}
e^{i\theta/2}(\boldsymbol{\sigma}\cdot\vex{d}')\sigma_{2}=\left(
\begin{array}{cc}
i e^{i\alpha} e^{i\theta} & 0\\
0 & -ie^{-i\alpha}\end{array}\right)
\end{equation}
Therefore, there is a vortex only in the spin-up component. Then, the zero energy solution is
\begin{equation}
\chi'_{1}=\frac{1}{\sqrt{2}}\left(
\begin{array}{c}
f v_\uparrow \\
f^{*} v_\uparrow
\end{array}
\right),
\quad\quad
\chi'_{2}=0.
\end{equation}
with an associated Majorana mode $\gamma'$. If we take y-axis to be at the intersection of the plane of rotation of $\vex d$ and the xy-plane, then $\hat{\vex N} \equiv \hat{\vex y}$, and the $\gamma'$ is found from Eq. (5) in the paper in the spin-down component with $S=e^{i\beta\sigma_y/2}$.

\section{The splitting of Majorana modes in the FQV}

Due to the presence of two spin components, two Majorana modes are present on a FQV. While these Majorana modes are protected against many types of local perturbations, they will split by the presence of perturbations that break spin rotation symmetry. These include the Zeeman splitting, $h_Z=\mu\,\boldsymbol{\sigma} \cdot {\vex B}$, and spin-orbit interactions, $h_{SO}=\boldsymbol{\lambda}_{so}\cdot (\boldsymbol{\sigma}\times \vex p)$.

We first consider the case that there are no external fields and no spin orbit interactions. When the $\vex d$ vector is directed along the y-axis, the BdG Hamiltonian decouples into the two spin components. One can then solve independently the BdG equation in each spin component, and then rotate back to the $\vex d$ direction using a rotation $S$
\begin{equation}\label{eq:rotatedMs}
	\chi_1=U\left(\begin{array}{c} f v_\uparrow \\ f^* v_\uparrow \end{array}\right), \quad\quad \chi_2=U\left(\begin{array}{c} f v_\downarrow \\ f^* v_\downarrow \end{array}\right).
\end{equation}
where $U$ is given by Eqs.~(\ref{eq:SU2}) and~(\ref{eq:SS}).
We now add the terms $H_Z$ and $H_{SO}$. The key point is the transformation properties of these objects under rotations. 

Starting from $H_Z$,
\begin{widetext}
\begin{equation}
	H_Z=\left(\begin{array}{cc}\mu\,\boldsymbol{\sigma} \cdot {\vex B} & 0 \\ 0 & -\mu(\boldsymbol{\sigma} \cdot {\vex B})^T \end{array}\right)\to \left(\begin{array}{cc}\mu\,S^\dag \boldsymbol{\sigma} \cdot {\vex B}S & 0 \\ 0 & -\mu(S^\dag \boldsymbol{\sigma} \cdot {\vex B} S)^T \end{array}\right)
\end{equation}
Since $S^\dag \boldsymbol{\sigma} \cdot {\vex B} S=\boldsymbol{\sigma} \cdot R_S{\vex B}$, $\vex B$ transforms as a vector. We can therefore take $\vex d$ to lie along the y-axis, and $\vex B$ in an arbitrary direction. We then get
\begin{equation}
	\lambda=-i\langle \chi_1|H_Z|\chi_2\rangle=\mu B_y=\mu \vex B\cdot \vex d
\end{equation}

The spin-orbit term when $\boldsymbol\lambda_{so} || \hat{\vex z}$ was considered in Ref.~\onlinecite{LuYip08a} and found not to lift the degeneracy of the two Majorana modes. Here we show that, to the leading order in perturbation theory, this is true for general $\boldsymbol\lambda_{so}$ and any direction of $\vex d$. Using our solution in Eq.~(\ref{eq:rotatedMs}) we have
\begin{equation}
H_{SO} = \left(
\begin{array}{cc}
\boldsymbol{\lambda_{so}}\cdot(\boldsymbol{\sigma}\times\vex p) & 0 \\
 0 & -\boldsymbol{\lambda_{so}}\cdot(\boldsymbol{\sigma}\times\vex p)^T 
\end{array}
\right)
\to 
\left(
\begin{array}{cc}
 S^\dag \boldsymbol{\lambda_{so}}\cdot(\boldsymbol{\sigma}\times\vex p) S & 0 \\
 0 & -(S^\dag \boldsymbol{\lambda_{so}}\cdot(\boldsymbol{\sigma}\times\vex p) S)^T 
\end{array}
\right).
\end{equation}
\end{widetext}
Since $S^\dag \boldsymbol{\lambda_{so}}\cdot(\boldsymbol{\sigma}\times\vex p) S=\boldsymbol{\sigma} \cdot R_S(\vex p\times\boldsymbol{\lambda_{so}})$, we can find the splitting in the basis where $\vex d$ is along the y-axis by rotating $\vex p$ and $\boldsymbol\lambda_{so}$ by angle $\delta$ around $\hat{\vex n}$. Therefore, in general the overlap $\langle \chi_1|H_{SO}|\chi_2\rangle$ will contain terms involving components of the integral $\int (f^*\vex p f) d\vex r$. Noting that $f$ and $f^*$ differ by a constant phase, we find $\int (f^* \vex p f)d\vex r \propto \int (\partial_{\vex r} |f|^2)d\vex r = 0$ since $f$ vanishes at infinity. We conclude
\begin{equation}
\langle \chi_1|H_{SO}|\chi_2\rangle = 0.
\end{equation}

\section{Effects of multiple layers in S\lowercase{r}$_2$R\lowercase{u}O$_4$}

\subsection{Non-Abelian statistics in the absence of inter-layer tunneling}

Sr$_2$RuO$_4$ is a layered material, and when a HQV is threaded through the system it goes through multiple layers. In the absence of tunneling between the layers, there will be one Majorana mode per layer bound to the vortex. Here we show that even in this limit the physics does not decouple into layers, but in fact results in an even-odd effect in the number of layers.

We denote the layer index by $\ell=1,\ldots,N$. We start with a HQV going around a second HQV. For this case, the Majorana modes on each of the layers for the two vortices acquire a minus sign. This is generated by the following transformation on the zero-energy Hilbert space,
\begin{equation}
	\label{eq:U-HQV}
	U'_{(\frac{1}{2},\frac{1}{2})}=\prod_{\ell=1}^N \gamma_\ell \gamma'_\ell.
\end{equation}
To explore the effect of this transformation on the interference term, we should understand whether such transformations for different itinerant vortices will commute~\cite{SteHal06a}, i.e. whether $[U',U'']=0$. It is easy to see that for $\ell$ even they commute, while for $\ell$ odd they do not commute, giving rise to the obliteration of the AC oscillations only in the latter case.

In the case that a FQV goes around a HQV, we denote by $\gamma'_1$ and $\gamma'_2$ the Majorana modes on the FQV, and by $\gamma_2$ the one on the HQV. The transformation $U'$ associated with the FQV encircling the HQV is given by
\begin{equation}
	\label{eq:HQV-FQV} U'_{(\frac{1}{2},1)}=\prod_{\ell=1}^N\gamma_{2,\ell} \gamma'_{2,\ell}
\end{equation}
which takes $\gamma'_{1,\ell}\to \gamma'_{1,\ell}$, but $\gamma_{2,\ell}\to -\gamma_{2,\ell}$ and $\gamma'_{2,\ell}\to -\gamma'_{2,\ell}$, being the same transformation (per layer) as in the main text. As the transformations (\ref{eq:U-HQV}) and (\ref{eq:HQV-FQV}) are the same, the results are the same: the obliteration of the AC oscillations will only occur for $N$ odd.

Finally, for a FQV encircling a FQV, the associated transformation is
\begin{equation}
	U'_{(1,1)}=\prod_{\ell=1}^N \gamma_{1,\ell}\gamma_{2,\ell}\gamma'_{1,\ell}\gamma'_{2,\ell}
\end{equation}
By relabeling the Majorana modes, the case of $N$ layers here is analogous to $2N$ layers for a HQV encircling a second HQV. Therefore, no effect on the interference pattern is expected in this case. 

To conclude, when either a HQV or a FQV encircles a HQV, we predict an even-odd effect in the number of layers: non-Abelian statistics will be present only in the case of an odd number of layers. In contrast, when a FQV encircles another FQV, the braiding statistics is abelian.

\subsection{Effects of inter-layer tunneling on the Majorana modes}

Inter-layer tunneling can be introduced between the Majorana states on a vortex (we consider a HQV for simplicity). We note that the tunneling of Majorana modes is a different physical phenomenon from that of Cooper pairs. For instance, when the sample has phase coherence along the z direction (i.e. in the absence of interlayer supercurrents) and contains a single chiral domain, the Majorana tunneling amplitude is exactly zero, even though Cooper pairs are free to move in all directions. This can be understood in two ways~\cite{Vol03a}.

First, since the z direction only appears in the kinetic energy, the momentum $p_z$ along the z-axis is a good quantum number. For a given $p_z$ the chemical potential is shifted as $\mu \to \mu - p_z^2/2m$.  So, there is a degenerate flat band of zero energy Majorana states in the weak pairing regime, $\mu - p_z^2/2m > 0$~\cite{LuYip08a}. Equivalently, from the solution to the Majorana modes above, we can see that the overlap amplitude of two Majoranas on adjacent layers is proportional to $\sin(\Delta\phi/2)$, where $\Delta\phi$ is the phase difference between the two layers. Phase coherence means $\Delta\phi=0$ and therefore all the tunneling amplitudes vanish.

This degeneracy is only protected by the additional symmetry in this idealized example and is lifted by phase fluctuations and other symmetry-breaking factors. But this example illustrates that the interlayer Majorana tunneling is subtle and quite different from Cooper pair tunneling.

Let us now examine the spectrum of Majorana modes when there is a finite interlayer Majorana tunneling. The tunneling term acquires the general form
\begin{equation}
	H=i \sum_{\ell=1}^{N-1}t_\ell\gamma_\ell \gamma_{\ell+1}
\end{equation}
We look for eigenstates of the form $\Gamma^\dag=\sum_{\ell=1}^N u_\ell \gamma_\ell$, satisfying $[H,\Gamma^\dag]=E\Gamma^\dag$. When $N$ is odd there is always a zero energy solution to this equation, as is clear by the particle-hole symmetry of the Hamiltonian. Writing the zero energy eigenvector of the Hamiltonian as $(u_1,\ldots,u_N)^T$, it is easy to see that $u_{2n}=0$ ($n\in \mathbb{N}$) and 
\begin{equation}
	u_{2n+1}=u_{2n-1}(t_{2n-1}/t_{2n})=u_1\prod_{n'=1}^{n}(t_{2n'-1}/t_{2n'}),
\end{equation}	
where $u_1$ is chosen by the requirement of normalization of the eigenvector. When all tunneling terms are non-zero, the Majorana fermion is spread through the odd numbered layers, avoiding the energy cost associated with nearest neighbors. When the number of layers is even, there are two such choices for states which then further hybridize and move away from zero energy. As we argued before, for an even number of layers we do not expect non-Abelian statistics to play a role in the AC interference.

When the number of layers is odd, a second quantity which becomes of interest to us is the width of the resulting band $W$. This width is set by the matrix elements $t_i$, $W\propto \max_i t_i$. At zero temperature, the width of the band is of negligible consequence for a proper choice of the Majorana fermion operator. At finite temperatures, de-coherence results due to the finite band, with a typical time-scale which goes to infinity as $W\to 0$ or $T\to 0$.



\end{document}